\documentclass[a4paper, 10pt]{report}

\usepackage[cp1250]{inputenc}
\usepackage[T1]{fontenc}
\usepackage{amsfonts}
\usepackage{graphicx}
\usepackage{amsmath}
\usepackage{caption}

\usepackage{amssymb}

\usepackage{fancyhdr}
\usepackage{bibtopic}
\usepackage{color}
\usepackage{ulem}
\usepackage[toc,page]{appendix}
\usepackage{setspace}
\usepackage{cite}

\AtBeginDocument{}
\usepackage{etoolbox}
\patchcmd{\thebibliography}{\chapter*}{\section*}{}{}

\makeatletter
\renewcommand{\thesection}{%
  \ifnum\c@chapter<1 \@arabic\c@section
  \else \thechapter.\@arabic\c@section
  \fi
}
\makeatother

\patchcmd{\tableofcontents}{\chapter*}{\section*}{}{}

\numberwithin{equation}{section}

\let\OLDthebibliography\thebibliography
\renewcommand\thebibliography[1]{
  \OLDthebibliography{#1}
  \setlength{\parskip}{0pt}
  \setlength{\itemsep}{3.5pt plus 1ex}
}

\usepackage[linktocpage]{hyperref}
\definecolor{darkred}{rgb}{0.5,0,0}
\definecolor{darkpurple}{rgb}{0.5,0,0.5}
\definecolor{darkblue}{rgb}{0,0,0.5}
\hypersetup{ colorlinks,
linkcolor=darkblue,
filecolor=darkpurple,
urlcolor=darkred,
citecolor=darkpurple }
 
\begin{document}

\allowdisplaybreaks
\setlength{\abovedisplayskip}{3.5pt}
\setlength{\belowdisplayskip}{3.5pt}
\abovedisplayshortskip
\belowdisplayshortskip

{\setstretch{1.0}

{\LARGE \bf \centerline{
Properties of generalized Schwarzschild spacetimes
}}
{\LARGE \bf \centerline{
with extra dimensions
}}

\vskip 1cm
\begin{center}
{Peter M\'esz\'aros\footnote{e-mail address: peter.meszaros@fmph.uniba.sk}}

\vskip 2mm {\it Department of Theoretical Physics, Comenius
University, Bratislava, Slovakia}

\vskip 2mm \today 
\end{center}

\section*{Abstract}

We show that an ansatz for $1+3+n$ dimensional static spacetime with spherical symmetry in three dimensions and Euclidean symmetry in $n$ dimensions, parametrized by only one function of radial coordinate, leads to a limited set of vacuum solutions of the Einstein field equations. They can also be identified as Weyl solutions. We investigate properties of these spacetimes through the Kretschmann scalar, Newtonian mass defined through the Newtonian limit, Komar mass, Einstein, Landau--Lifshitz, and ADM mass. In addition to $1+3+n$ dimensional Minkowski spacetime, there are two classes of solutions. The first class is a trivial product of the Schwarzschild spacetime and Euclidean spaces in extra dimensions, while the second class is non-trivial. In the case with no horizon, there is a naked singularity, all masses are equal, and they are negative. In the case when there is a horizon, this horizon accommodates a physical singularity, which corresponds to Kaluza--Klein bubbles featuring exotic properties. Einstein, Landau--Lifshitz, and ADM masses are positive, while Newtonian and Komar masses are negative. This differentiates these solutions from trivial higher-dimensional extensions of the Schwarzschild solution.

\vskip 3mm \hspace{4mm}
\begin{minipage}[t]{0.8\textwidth}
\noindent\rule{12cm}{0.4pt}
\vspace{-9mm}
\tableofcontents
\noindent\rule{12cm}{0.4pt}
\end{minipage}
\vskip 6mm

}

\section{Introduction}\label{sec:1}

Schwarzschild spacetime \cite{schw} is one of the most prominent examples of vacuum solutions of the Einstein field equations. It can be interpreted as a black hole \cite{finkelstein}, and such objects are present in our Universe \cite{black1,black2,black3}. Black holes generate Hawking radiation \cite{hawking}, providing a laboratory for testing quantum effects in gravity. Modification of the original Schwarzschild solution to an arbitrary number of space dimensions was first found by Tangherlini \cite{tanger}. Rotating black holes with the Kerr metric \cite{kerr} can also be generalized to an arbitrary number of space dimensions. They are known as Myers--Perry black holes \cite{mp1,mp2}. There are also other different higher-dimensional generalizations \cite{emparan}, including, for example, black rings \cite{ring}.

\vskip 2mm
Physical motivation for extra dimensions originates mainly from Kaluza--Klein models \cite{kaluza,kal1} attempting to solve the hierarchy problem \cite{kal2,kal3} and string theory \cite{witten,zwiebach,becker} developed as a promising candidate for a quantum theory of gravity. These theories usually require some kind of compactification of extra dimensions. Small extra dimensions can be present in a different class of objects known as black strings, p-branes, or black branes \cite{horowitz,kanti,brecher,lunin,faril,orlov}. They may have black hole properties in macroscopic dimensions while containing compactified extra dimensions from String theory \cite{string1,string2,string3,string4}. Higher-dimensional objects provide valuable testing grounds for studying quantum effects in general relativity \cite{rad1,rad2,rad3}.

\vskip 2mm
This paper provides an analysis of a special class of vacuum solutions in $1+3+n$ dimensions under the assumption of spherical symmetry in the three-dimensional part and Euclidean symmetry in the $n$-dimensional part, $SO(3)\times E(n)$. Extra dimensions may be compactified into a flat $n$-torus with $S^1\times...\times S^1$ topology. The number of the standard space dimensions that are not compactified is fixed to three in this work since it is the physically most relevant case. Disregarding $1+3+n$ dimensional Minkowski spacetimes, there will be two different classes of solutions. The first class represents a simple extension of the $1+3$ dimensional Schwarzschild spacetime with the addition of $n$ flat dimensions, with $g_{AB}=\delta_{AB}$, where indices $A$ and $B$ run through $n$ additional dimensions. In contrast, the second class features a non-trivial form of $g_{AB}$ components of the spacetime metric. These solutions can also be found through Weyl solutions \cite{weyl} and their higher-dimensional generalizations \cite{emparangen} with the non-trivial class corresponding to so-called Kaluza-Klein bubbles \cite{witten0,elvang}.

\vskip 2mm
The following section \ref{sec:2} provides vacuum solutions of the Einstein field equations with two classes mentioned above, section \ref{sec:3} includes further analysis of properties of these spacetimes based on the Kretschmann scalar, Newtonian limit, Komar mass, conserved masses from Einstein and Landau--Lifshitz pseudotensors, and ADM mass. Section \ref{sec:4} is dedicated to summary and discussion. The usual convention with the light speed set to unity, $c=1$, together with the spacetime metric signature $(-,+,+,...)$ will be used throughout this paper. The index $0$ will be used for a time coordinate, lowercase Latin indices will indicate three space coordinates, $i=1,2,3$, and capital Latin indices will run through $n$ extra dimensions, $A=1,...,n$. All spacetime indices will be collectively denoted by Greek letters, $\mu=0, i, A$.

\section{Two classes of vacuum solutions}\label{sec:2}

The existence of two different classes of extensions of the Schwarzschild spacetime is obvious in the case with only one extra space dimension. The metric of the trivial case can be written as
\begin{eqnarray}\label{eq:s1}
ds^2 = -\left(1+\dfrac{a}{r}\right) dt^2 + \left(1+\dfrac{a}{r}\right)^{-1} dr^2 + r^2 d\Omega_{(2)}^2 + d\zeta^2,
\end{eqnarray}
where $t$ is time, $a$ is a constant, $r$ is radial coordinate, $d\Omega_{(2)}^2=d\vartheta^2+\sin^2\vartheta d\varphi^2$ is line element of a unit $2$-sphere, and $\zeta$ denotes coordinate in the additional dimension. Since the extra dimension is added most trivially, the spacetime with this metric is a vacuum solution like the original Schwarzschild spacetime. The Ricci tensor remains zero if this metric is rewritten through a formal change $\zeta\to it$, $t\to i\zeta$. The resulting metric reads
\begin{eqnarray}\label{eq:s2}
ds^2 = -dt^2 + \left(1+\dfrac{a}{r}\right)^{-1} dr^2 + r^2 d\Omega_{(2)}^2 + \left(1+\dfrac{a}{r}\right) d\zeta^2.
\end{eqnarray}
Metrics (\ref{eq:s1}) and (\ref{eq:s2}) describe two spacetimes with different properties, which, for example, can be tested by a slowly moving particle far away from the center in the Newtonian limit. Metric (\ref{eq:s1}) represents the simplest black string or black ring, and (\ref{eq:s2}) is the simplest Kaluza--Klein bubble. This simple example with only one extra space dimension can be generalized to an arbitrary number of extra dimensions, which will be examined in what follows.

\vskip 2mm
Our goal is to find a static vacuum solution of the Einstein field equations with three-dimensional spherical symmetry and $n$-dimensional Euclidean symmetry in extra dimensions, $SO(3)\times E(n)$. This can also be interpreted as a stack of uniform $n$-branes in $1+3+n$ dimensions. We start with the ansatz
\begin{eqnarray}\label{eq:ansatz}
ds^2 = -f^{\alpha}dt^2+f^{\beta}dr^2+r^2 d\Omega_{(2)}^2+f^{\gamma}\delta_{AB}d\zeta^A d\zeta^B,
\end{eqnarray}
where $\alpha$, $\beta$ and $\gamma$ are constants, and $f=f(r)$ is a function of only the radial coordinate. A more general ansatz may be parametrized by two \cite{astro} or three independent functions of the radial coordinate \cite{lunin}, but in our analysis, we assume that these functions are given by different powers of one function $f(r)$.

\vskip 2mm
There are two reasons to start with the special choice of ansatz (\ref{eq:ansatz}). A pragmatic reason is that with this choice, the condition of the Ricci tensor being zero reduces to simple algebraic relations. The second reason is that one would expect only one potential event horizon, which occurs only when all components of the spacetime metric $g_{00}$, $g_{rr}$ and $g_{AB}$ are singular at the same radial coordinate, and the simplest way to achieve this is to consider our form of the ansatz. In principle, there may be vacuum solutions with $SO(3)\times E(n)$ symmetry exhibiting more than one horizon; however, with the ansatz (\ref{eq:ansatz}), we will not be able to find them. 

\vskip 2mm
All nonzero components of the Ricci tensor in spacetime with the metric of the form (\ref{eq:ansatz}) can be written as
\begin{eqnarray}
& & R_{00}=\alpha f^{\alpha-\beta}F_1, \quad
R_{rr}=F_2, \quad
R_{\vartheta\vartheta}=F_3, \\
& & R_{\varphi\varphi}=F_3\sin^2\vartheta, \quad
R_{AB}=-\gamma f^{\gamma-\beta}F_1\delta_{AB}, \nonumber
\end{eqnarray}
where
\begin{eqnarray}\label{eq:Fs}
& & F_1 = \dfrac{1}{r}\dfrac{f^{\prime}}{f} + \dfrac{1}{4}\left(\alpha-\beta+n\gamma-2\right)\left(\dfrac{f^{\prime}}{f}\right)^2 + \dfrac{1}{2}\dfrac{f^{\prime\prime}}{f}, \\
& & F_2 = \beta\dfrac{1}{r}\dfrac{f^{\prime}}{f} + \dfrac{1}{4}\left[\alpha\left(-\alpha+\beta+2\right)+n\gamma(\beta-\gamma+2)\right]\left(\dfrac{f^{\prime}}{f}\right)^2 - \dfrac{1}{2}\left(\alpha+n\gamma\right)\dfrac{f^{\prime\prime}}{f}, \nonumber\\
& & F_3 = 1 - f^{-\beta}\left[1+\dfrac{1}{2}\left(\alpha-\beta+n\gamma\right)r\dfrac{f^{\prime}}{f}\right], \nonumber
\end{eqnarray}
with prime denoting differentiation with respect to the radial coordinate. All three functions $F_1$, $F_2$, and $F_3$ must be zero to have the vacuum solution.

\vskip 2mm
The condition $F_3=0$ represents a simple differential equation that can be solved by integration. The result reads
\begin{eqnarray}\label{eq:form}
f = \left( 1 + a r^q \right)^{-1/\beta}, \quad q = \dfrac{2\beta}{\alpha-\beta+n\gamma},
\end{eqnarray}
where $a$ is an integration constant, and $q$ is given by constants from the ansatz. This solution is invalid if $\alpha-\beta+n\gamma=0$, and this case has to be treated separately. It is easy to see that in this case, the condition $F_3=0$ implies $f=1$, and we retrieve the simple $1+3+n$ dimensional Minkowski spacetime. The same is true if the integration constant $a$ is zero. By inserting (\ref{eq:form}) into expressions for $F_1$ and $F_2$ in (\ref{eq:Fs}) we find
\begin{eqnarray}
F_1 = C_1\Phi, \quad F_2 = (C_2+C_3ar^q)\Phi,
\end{eqnarray}
where function $\Phi$ and constants $C_1$, $C_2$ and $C_3$ are
\begin{eqnarray}
& & \Phi = \dfrac{1}{\left(\alpha-\beta+n\gamma\right)^2}\dfrac{ar^{q-2}}{\left(1+ar^q\right)^2}, \\
& & C_1 = -\alpha-\beta-n\gamma, \nonumber\\
& & C_2 = 2\beta^2+\beta\left(\alpha+n\gamma\right)-\left(\alpha+n\gamma\right)^2, \nonumber\\
& & C_3 = 2\beta^2-\alpha^2-n\gamma^2-\left(\alpha+n\gamma\right)^2. \nonumber
\end{eqnarray}
The problem of finding vacuum solutions of the form (\ref{eq:ansatz}) is now reduced to the system of algebraic equations, $C_1=C_2=C_3=0$.

\vskip 2mm
First two conditions $C_1=0$ and $C_2=0$ are satisfied if $\alpha=-\beta-n\gamma$, which implies $q=-1$, and by inserting this expression for $\alpha$ into the relation for $C_3$, we also find
\begin{eqnarray}
C_3 = -n\gamma\left[2\beta+\left(n+1\right)\gamma\right].
\end{eqnarray}
The condition $C_3=0$ is then satisfied by two sets of possible solutions. One set corresponds to the choice $\gamma=0$, and for the second set, we have $2\beta+(n+1)\gamma=0$. With the choice $\gamma=0$ we have $\alpha=-\beta$. Components of the spacetime metric are then given by $f^{\alpha}=1+ar^{-1}$, $f^{\beta}=\left(1+ar^{-1}\right)^{-1}$, and $f^{\gamma}=1$, and the corresponding spacetime metric can be written as
\begin{eqnarray}\label{eq:trivial}
ds^2 = -\left(1+\dfrac{a}{r}\right)dt^2 + \left(1+\dfrac{a}{r}\right)^{-1}dr^2 + r^2 d\Omega_{(2)}^2 + \delta_{AB}d\zeta^Ad\zeta^B.
\end{eqnarray}
This is only a trivial extension of the original $1+3$ dimensional Schwarzschild solution, with metric (\ref{eq:s1}) as a special case for $n=1$. More interesting solutions correspond to the second choice with $2\beta+(n+1)\gamma=0$, where $\alpha/\beta=(n-1)/(n+1)$, and $\gamma/\beta=-2/(n+1)$. Spacetime metric then can be written as
\begin{eqnarray}\label{eq:metric}
ds^2 = -\left(1+\dfrac{a}{r}\right)^{-\frac{n-1}{n+1}}dt^2 + \left(1+\dfrac{a}{r}\right)^{-1}dr^2 + r^2 d\Omega_{(2)}^2 + \left(1+\dfrac{a}{r}\right)^{\frac{2}{n+1}}\delta_{AB}d\zeta^Ad\zeta^B.
\end{eqnarray}
We can also check that for $n=1$ we retrieve (\ref{eq:s2}). In the case with no extra dimensions, $n=0$, both sets of solutions (\ref{eq:trivial}) and (\ref{eq:metric}) automatically reduce to the Schwarzschild solution. From now on, we will address (\ref{eq:trivial}) as the trivial case and (\ref{eq:metric}) as the non-trivial. Solutions (\ref{eq:trivial}) and (\ref{eq:metric}) differ in $g_{00}$ and $g_{AB}$ components. The form of $g_{00}$ determines gravitational redshift as well as the way of fixing the integration constant $a$ by relating it to the gravitating mass in the Newtonian limit $M_{\textrm{New.}}$, while $g_{AB}$ components measure the size of extra dimensions. This will be discussed in more detail in the next section.

\vskip 2mm
Spacetime metric in the non-trivial case (\ref{eq:metric}) can be written also in the generalized Weyl form \cite{emparangen},
\begin{eqnarray}\label{eq:weyl}
ds^2 = -e^{2A}dt^2 + e^{2B}d\varphi^2 + e^{2C}\delta_{AB}d\zeta^A d\zeta^B + e^{2D}\left(d\rho^2+dz^2\right),
\end{eqnarray}
where functions $A$, $B$, $C$, and $D$ depend on two space coordinates $\rho$ and $z$, and they are solutions of Laplace's equation in cylindrical coordinates $(\rho,\varphi,z)$ with infinitely thin rod segments representing sources at $\rho=0$. By choosing these functions as
\begin{eqnarray}
& & A=\frac{1}{2}\frac{1-n}{1+n}\ln\frac{Q+a}{Q-a}, \quad B=\ln\rho-\frac{1}{2}\ln\frac{Q+a}{Q-a}, \\
& & C=\frac{1}{1+n}\ln\frac{Q+a}{Q-a}, \quad D=\frac{1}{2}\ln\frac{(L-a)^2}{4Q_{+}Q_{-}}, \nonumber
\end{eqnarray}
where $Q=Q_{+}+Q_{-}$, $Q_{\pm}=\sqrt{(z\pm a/2)^2+\rho^2}$, and by performing the coordinate transformation $z=r(1+a/(2r))\cos\vartheta$, $\rho=r\sqrt{1+a/r}\sin\vartheta$, we can rewrite (\ref{eq:weyl}) into the form (\ref{eq:metric}).

\vskip 2mm
Different solutions similar to (\ref{eq:metric}) can be found in the literature. Authors of \cite{horowitz} worked with the low-energy string theory action in ten dimensions, and they found a brane solution with a single Ramond--Ramond charge with non-trivial $g_{AB}$ components in terms of our notations. This has been studied further in many other works \cite{gregory,halyo,polchinski,ahn,lake,mourad}. Extra dimensions parametrized by coordinates $\zeta^A$ in (\ref{eq:metric}) may be macroscopic with $\mathbb{R}^n$ topology, microscopic with $n$-torus topology, or a combination of both cases with some macroscopic and some microscopic dimensions. In all cases, there is Euclidean symmetry in extra dimensions, but one can assign extra dimensions to the part that is only spherically symmetric \cite{felice}. Note also that considering extra time dimensions instead of extra space dimensions, or having mixed time and space extra dimensions, would not change the results because changing the signs of $g_{AB}$ components of the metric keeps the Ricci tensor zero. Compactification of the time dimension may also be considered, with implications for branes \cite{sugimoto}.

\vskip 2mm
We will study some properties of the found solutions (\ref{eq:trivial}) and (\ref{eq:metric}) in the following section. To focus on physically most relevant cases, let us assume that all extra dimensions are microscopic with $S^1\times...\times S^1$ topology of a flat $n$-torus. Without loss of generality, we may consider all extra coordinates $\zeta^A$ to be in the interval $\zeta^A\in[0,L)$, with the same periodicity $L$ for every $\zeta^A$.

\section{Physical properties}\label{sec:3}

In the case of the ordinary Schwarzschild metric, there is no physical singularity on the horizon. This can be easily revealed by the calculation of the Kretschmann scalar invariant defined through the Riemann curvature tensor as $K=R^{\mu\nu\lambda\sigma}R_{\mu\nu\lambda\sigma}$. For the Schwarzschild spacetime, it is $K_{\textrm{Schw.}}=12a^2r^{-6}$, indicating a singularity only at the origin, and the same is true also for the trivial extension (\ref{eq:trivial}). In the non-trivial case (\ref{eq:metric}) we have
\begin{eqnarray}\label{eq:kr0}
K = \dfrac{12a^2}{r^6}\left[1-\dfrac{1}{3}\dfrac{n(n-1)}{(n+1)^3}\left(1+\dfrac{r}{a}\right)^{-2}\left(\dfrac{3}{4}n+1+(n+1)\dfrac{r}{a}\right)\right].
\end{eqnarray}
We can see that for $n=0$ and $n=1$, there is no difference from the Schwarzschild spacetime, $K=K_{\textrm{Schw.}}$. Unless $a=0$, there is always the central singularity at $r=0$. In the case with positive $a$, there is no additional singularity, but for negative values of constant $a$, the Kretschmann scalar indicates a singularity at the horizon, $r=-a$, for $n\geq 2$. Other examples of such singularities can be found, for example, in \cite{san,pul}.

\vskip 2mm
Since there is a horizon which is not associated with any physical singularity in the Schwarzschild solution, it is possible to find its maximal analytic extension \cite{kruskal}. Obviously, the same is true also for the trivial case with extra dimensions (\ref{eq:trivial}) if $a$ is negative. In other cases, there is no possibility to construct the maximal extension either because of the lack of horizon if $a$ is positive, or because of the presence of the physical singularity at the horizon. This also means that two regions inside and outside the spherical singularity represent two separate spacetimes. The spacetime corresponding to the outer region has properties of a Kaluza--Klein bubble \cite{witten0,elvang}, since from the point of view of the outer region, there is no spacetime below the spherical singularity.

\vskip 2mm
The Newtonian limit provides a very important insight into properties of branes containing spherically symmetric $1+3$ dimensional part \cite{cot,iva}. We will examine solutions (\ref{eq:trivial}) and (\ref{eq:metric}) in this limit under the assumption of $L$-periodicity of coordinates $\zeta^A$ in extra dimensions. Macroscopic objects of size larger than $L$ can move only in the $1+3$ dimensional section. The same is true for sufficiently long-wavelength photons or particles with large enough de Broglie wavelengths. From the perspective of such objects, the spacetime in regions far from the origin, $r\gg |a|$, can be approximated as
\begin{eqnarray}
ds^2 \approx -(1+2\phi) dt^2 + (1-2\psi)\delta_{ij}dx^idx^j,
\end{eqnarray}
where $\phi$ is the Newtonian potential. Without any extra microscopic dimensions, it can be shown that $\psi=\phi$ in harmonic coordinates. The geodesic equation for a slowly moving object reduces to the Newtonian equation of motion, $\ddot{x}^i \approx -\phi_{,i}$, in this limit.

\vskip 2mm
In order to examine the Newtonian limit for spacetimes with metrics (\ref{eq:trivial}) and (\ref{eq:metric}), we have to replace the radial coordinate $r$ with the coordinate $\rho$ defined by the transformation $r = \rho\left(1-a/(4\rho)\right)^2$. This new coordinate is usually called the isotropic radial coordinate because in the case of $1+3$ dimensional Schwarzschild metric, the space part of the metric becomes conformally Euclidean. In our case, this is true only for the three-dimensional and $n$-dimensional sections separately. For positive values of $a$, the isotropic radial coordinate $\rho$ covers the entire spacetime except for the central singularity twice, while for negative $a$, it covers twice only the region above the horizon. To avoid the double coverage, we may consider $\rho$ only in the interval $[|a|/4,\infty)$. In the trivial case (\ref{eq:trivial}) we have
\begin{eqnarray}
ds^2 = - \left(\dfrac{\chi_{+}}{\chi_{-}}\right)^2 dt^2 + \left(\chi_{-}\right)^4 \delta_{ij}dx^idx^j + \delta_{AB}d\zeta^Ad\zeta^B,
\end{eqnarray}
where functions $\chi_{+}$ and $\chi_{-}$ are defined as $\chi_{\pm}=1 \pm a/(4\rho)$, and space coordinates $x^i$ are Cartesian such that $(x^1)^2+(x^2)^2+(x^3)^2=\rho^2$. This leads to
\begin{eqnarray}\label{eq:limtrivial}
ds^2 \approx - \left( 1 + \dfrac{a}{\rho} \right) dt^2 + \left( 1 - \dfrac{a}{\rho} \right) \delta_{ij} dx^i dx^j + \delta_{AB} d\zeta^A d\zeta^B,
\end{eqnarray}
in the Newtonian limit, $\rho\gg|a|$, with the form of the Newtonian potential, $\phi=(a/2)\rho^{-1}$, and the usual relation between the constant $a$ and the mass of the gravitating body in the Newtonian theory, $a=-2\kappa M_{\textrm{New.}}$, where $\kappa$ denotes the Newtonian gravitation constant. In the non-trivial case (\ref{eq:metric}), the spacetime metric can be rewritten as
\begin{eqnarray}\label{eq:limmetric}
& & ds^2 = - \left(\dfrac{\chi_{+}}{\chi_{-}}\right)^{-2\frac{n-1}{n+1}} dt^2 + \left(\chi_{-}\right)^4 \delta_{ij}dx^idx^j + \left(\dfrac{\chi_{+}}{\chi_{-}}\right)^{\frac{4}{n+1}} \delta_{AB}d\zeta^Ad\zeta^B \approx \\
& & \phantom{ds^2} \approx - \left( 1 - \dfrac{n-1}{n+1} \dfrac{a}{\rho} \right) dt^2 + \left( 1 - \dfrac{a}{\rho} \right) \delta_{ij} dx^i dx^j + \left( 1 + \dfrac{2}{n+1} \dfrac{a}{\rho} \right) \delta_{AB} d\zeta^A d\zeta^B \nonumber,
\end{eqnarray}
and the Newtonian mass is related to the constant $a$ through a different relation for the Newtonian potential
\begin{eqnarray}\label{eq:Mnew}
\phi = -\dfrac{n-1}{n+1}\dfrac{a}{2}\dfrac{1}{\rho} = -\kappa M_{\textrm{New.}} \dfrac{1}{\rho}.
\end{eqnarray}
A consequence of a negative value of $a$ is the existence of a horizon. However, in the modified case with the spacetime metric (\ref{eq:metric}), the sign in the relation above is opposite, which implies naked singularities for positive masses $M_{\textrm{New.}}$. Negative $a$ and the existence of a horizon is then associated with negative mass $M_{\textrm{New.}}$, which means repulsive gravity in the Newtonian limit. Other examples of different types of black holes with such a property are studied, for example, in \cite{luongo}. We may also notice that the presence of extra dimensions does not change the position of the horizon or its properties considerably. For example, a contrary case can be found in \cite{silva}. Note also that for $n=1$, the spacetime metric in the non-trivial case is (\ref{eq:s2}), and the Newtonian potential vanishes, which means zero Newtonian mass $M_{\textrm{New.}}$.

\vskip 2mm
The physical length of periodicity in extra dimensions is proportional to $p_{\textrm{phys.}} \propto \sqrt{g_{AB}}$. We can see that the size of extra dimensions is constant in entire spacetime in the trivial case (\ref{eq:trivial}), while in the non-trivial case (\ref{eq:metric}), it depends on the radial coordinate, collapsing to zero size on the horizon if $a$ is negative and also diverging to infinity at the central singularity regardless of the sign of $a$. Let us assume that the maximal size of extra dimensions is at the Planck distance $l_{\textrm{Pl}}$ from the center, since one cannot approach any point at a closer distance. By calculating the physical radial distance as $r_{\textrm{phys.}} = \int_0^{r} \sqrt{g_{rr}} dr$ and taking the limit $r\ll a$, we find the ratio of physical periodicity at the Planck distance from the center to its value very far from the origin,
\begin{eqnarray}
\lambda=\dfrac{p_{\textrm{phys.}}(r_{\textrm{phys.}}=l_{\textrm{Pl}})}{p_{\textrm{phys.}}(r_{\textrm{phys.}}\to\infty)} \approx \left[ 1.22\cdot 10^{38} \dfrac{n+1}{n-1} \dfrac{M_{\textrm{New.}}}{M_{\textrm{sol.}}} \right]^{\frac{2}{3}\frac{1}{n+1}},
\end{eqnarray}
where we have used also (\ref{eq:Mnew}), and $M_{\textrm{sol.}}$ denotes the solar mass. For supermassive black holes with a mass of the order $10^{10} M_{\textrm{sol.}}$ and the minimal number of extra dimensions allowing horizon singularity $n=2$, we have $\lambda \approx 6\cdot 10^{10}$, and this factor becomes smaller for smaller masses as well as for higher numbers of extra dimensions. This means that if the size of extra dimensions is microscopic far from the origin, it remains very small even near the central singularity.

\vskip 2mm
In addition to the Newtonian mass, there are also other ways to assign a mass-like quantity to spacetimes or spacetime regions. Komar mass \cite{komar} associated with space region $\Omega$ can be calculated as \cite{cohen,kulkarni}
\begin{eqnarray}
M_{\textrm{Kom.}} = - \dfrac{1}{8\pi\kappa} \int\limits_{\partial\Omega} \star d\sigma,
\end{eqnarray}
where $\star d\sigma$ is dual of two form $d\sigma$ defined through timelike Killing one form $\sigma$. In the case with spacetime metric of the form (\ref{eq:trivial}) as well as (\ref{eq:metric}), we have $\sigma=g_{00}dt$, and $\star d\sigma = g_{00,r} d\vartheta \wedge d\varphi \wedge d\zeta^1 \wedge ... \wedge d\zeta^n$. By choosing the volume region $\Omega$ as a sphere in the three-dimensional space part with radius $r=R$, so that the topology of its boundary $\partial\Omega$ is $S^2 \times S^1 \times ... \times S^1$, we find
\begin{eqnarray}
M_{\textrm{Kom.}}(R) = -\dfrac{1}{2\kappa}R^2 g_{00,r}(R) L^n,
\end{eqnarray}
where $L^n=\int d\zeta^1...d\zeta^n$ is $n$-dimensional volume of the extra dimensions far from the origin, $r\to\infty$, where $g_{AB}\to\delta_{AB}$. By taking the limit $R\to\infty$, we find that the Komar mass coincides with the Newtonian mass discussed above up to the $n$-dimensional volume factor, $M_{\textrm{Kom.}}=M_{\textrm{New.}}L^n$.

\vskip 2mm
Another approach to defining conserved quantities characterizing properties of spacetimes is based on pseudotensors. Here, we consider Einstein \cite{einstein} and Landau--Lifshitz \cite{ll} stress-energy pseudotensors. Mass defined through the Einstein stress-energy pseudotensor can be calculated as \cite{moller}
\begin{eqnarray}
& & M_{\textrm{Ein.}} = \int\limits_{\partial\Omega} E_{0}^{\phantom{0}0i} dS_i, \\
& & E_{\mu}^{\phantom{\mu}\nu\rho} = \dfrac{1}{16\pi\kappa}\dfrac{g_{\mu\sigma}}{\sqrt{-g}} \left[ (-g)\left(g^{\nu\sigma}g^{\rho\lambda}-g^{\rho\sigma}g^{\nu\lambda}\right)\right]_{,\lambda}. \nonumber
\end{eqnarray}
Similarly, the mass defined through Landau--Lifshitz stress-energy pseudotensor is given by
\begin{eqnarray}
& & M_{\textrm{L.,L.}} = \int\limits_{\partial\Omega} L^{00i} dS_i \\
& & L^{\mu\nu\rho} = \dfrac{1}{16\pi\kappa} \left[ (-g) \left(g^{\mu\nu}g^{\rho\sigma}-g^{\mu\rho}g^{\nu\sigma}\right) \right]_{,\sigma}. \nonumber
\end{eqnarray}
By considering the same region $\partial\Omega$ as previously and by taking the limit $R\to\infty$, we find that $M_{\textrm{Ein.}}=M_{\textrm{L.,L.}}=-aL^n/(2\kappa)$ for the trivial case (\ref{eq:trivial}) with the use of its form in Newtonian limit (\ref{eq:limtrivial}). These two masses also equal the Komar mass as well as the Newtonian mass up to the factor $L^n$. On the other hand, in the non-trivial case with the metric of the form (\ref{eq:limmetric}), we obtain
\begin{eqnarray}
M_{\textrm{Ein.}} = M_{\textrm{L.,L.}} = - \dfrac{1}{n+1} \dfrac{a}{2\kappa}L^n.
\end{eqnarray}
Einstein and Landau--Lifshitz masses coincide, but they differ from Komar and Newtonian masses in this case.

\vskip 2mm
Finally, the ADM mass \cite{adm1,adm2,adm3} can be calculated as \cite{adm4}
\begin{eqnarray}
E = \dfrac{1}{16\pi\kappa} \oint\limits_{\partial\Omega} \sqrt{\gamma} \gamma^{\mu\nu}\gamma^{\alpha i} \left( \gamma_{\mu\alpha,\nu} - \gamma_{\mu\nu,\alpha} \right) d S_i,
\end{eqnarray}
where $\gamma_{\mu\nu}$ is metric of chosen spacelike hypersurface $\partial\Omega$. We choose it as in the previous cases. For the Schwarzschild spacetime with no extra dimensions, we have $E(R)=-(R/\kappa)\sqrt{1+a/R}$, and to obtain a meaningful result, we have to subtract the same quantity evaluated in the Minkowski spacetime with $a$ set to zero, $E_0(R)=-R/\kappa$. The total ADM mass then can be found as $M_{\textrm{ADM}} = \lim\limits_{R\to \infty}\left(E(R)-E_0(R)\right)=-a/(2\kappa)$. In the trivial case with extra compact dimensions (\ref{eq:trivial}), we have an expected result,
\begin{eqnarray}
M_{\textrm{ADM}} = \lim\limits_{R\to \infty} \left(E(R)-E_0(R)\right) = \lim\limits_{R\to \infty} \dfrac{R}{\kappa}\left(1-\sqrt{1+\dfrac{a}{R}}\right) L^n = -\dfrac{a}{2\kappa} L^n,
\end{eqnarray}
while in the non-trivial case with (\ref{eq:metric}), we find
\begin{eqnarray}
M_{\textrm{ADM}} & = &\lim\limits_{R\to \infty} \left(E(R)-E_0(R)\right) = \\
& = & \lim\limits_{R\to \infty} \dfrac{R}{\kappa}\left[1-\left(1+\dfrac{1}{2}\dfrac{n+2}{n+1}\dfrac{a}{R}\right)\left(1+\dfrac{a}{R}\right)^{\dfrac{1}{2}\dfrac{n-1}{n+1}}\right] L^n = -\dfrac{2n+1}{n+1} \dfrac{a}{2\kappa} L^n. \nonumber
\end{eqnarray}
The ADM mass is the same as other masses in the trivial case, while it differs from all of them in the non-trivial case.

\vskip 2mm
All masses calculated for the non-trivial case reduce to results for the trivial case by setting $n=0$. Therefore, all results can be summarized as
\begin{eqnarray}\label{eq:masses}
& & M_{\textrm{New.}} L^n = M_{\textrm{Kom.}} = \dfrac{n-1}{n+1} \dfrac{a}{2\kappa} L^n, \\
& & M_{\textrm{Ein.}} = M_{\textrm{L.,L.}} = - \dfrac{1}{n+1} \dfrac{a}{2\kappa} L^n, \quad M_{\textrm{ADM}} = - \dfrac{2n+1}{n+1} \dfrac{a}{2\kappa} L^n, \nonumber
\end{eqnarray}
with $n=0$ corresponding both to the original Schwarzschild spacetime as well as to its trivial extension with an arbitrary number of extra dimensions (\ref{eq:trivial}). We can see that signs of $M_{\textrm{Ein.}}$, $M_{\textrm{L.,L.}}$ and $M_{\textrm{ADM}}$ are opposite to signs of $M_{\textrm{New.}}$ and $M_{\textrm{Kom.}}$ for $n \geq 2$. Negative ADM mass of Kaluza--Klein bubbles has been found, for example, in \cite{negative}.

\section{Discussion}\label{sec:4}

This paper has provided an analysis of possible extensions of the Schwarzschild spacetime with $n$ extra dimensions and $SO(3) \times E(n)$ symmetry. There are three distinctive classes of static vacuum solutions of the Einstein field equations with metric given by the ansatz (\ref{eq:ansatz}). For $f=1$, we have $1+3+n$ dimensional Minkowski spacetime, and more interesting cases are given by metrics (\ref{eq:trivial}) and (\ref{eq:metric}). The first is a direct product of the $1+3$ dimensional Schwarzschild spacetime and $n$-dimensional Euclidean space. If we assume microscopic extra dimensions with $n$-torus topology, such spacetimes do not differ from the ordinary Schwarzschild spacetime from the point of view of physical objects larger than the size of extra dimensions. Most intriguing solutions are given by the metric (\ref{eq:metric}), and in the previous section, we investigated their nonstandard properties.

\vskip 2mm
For the number of extra dimensions $n \geq 2$, the Kretschmann scalar (\ref{eq:kr0}) of the spacetime metric (\ref{eq:metric}) diverges on the horizon, $r=-a$, which indicates a physical singularity on the horizon. Extra dimensions collapse on this horizon, which is a feature of Kaluza--Klein bubbles. There are three different masses in this case (\ref{eq:masses}). Newtonian mass and Komar mass coincide, $M_{\textrm{Kom.}}=M_{\textrm{New.}}L^n$, and Einstein, Landau--Lifshitz and ADM masses are given by
\begin{eqnarray}
M_{\textrm{Ein.}} = M_{\textrm{L.,L.}} = -\dfrac{1}{n-1} M_{\textrm{Kom.}},
\quad
M_{\textrm{ADM}} = -\dfrac{2n+1}{n-1} M_{\textrm{Kom.}}.
\end{eqnarray}
Existence of a physical singularity on the horizon is then associated with positive Einstein, Landau--Lifshitz, and ADM masses, and negative Newtonian and Komar masses. Three different masses originate from three independent components of the metric (\ref{eq:metric}). The Newtonian and Komar masses are given by only $g_{00}$ component, Einstein and Landau--Lifshitz masses have been calculated through the determinant $(-g)$, $g_{00}$ and $g_{ij}$ components, and to calculate the ADM mass we have used $g_{rr}$ and $g_{AB}$ components. A similar example is a class of spherically symmetric spacetimes with three space dimensions and with only two independent components that yield only two independent masses \cite{vollick}. In the trivial case (\ref{eq:trivial}) with a positive constant $a$, there is no horizon, and all masses discussed above are equal, and they are negative.

\vskip 2mm
Since different masses (\ref{eq:masses}) have different signs indicating nonphysical properties, we should also expect problems with instability. Some higher-dimensional extensions of the Schwarzschild spacetime tend to be unstable \cite{stab}, which also includes Kaluza--Klein bubbles \cite{witten0} and brane solutions \cite{stab1,stab2}. Cases with a positive constant $a$ feature a naked singularity, which also hints at potential problems with stability \cite{inst1,inst2}. The spacetime metric (\ref{eq:metric}) then represents a case in which different masses are associated with problematic properties.

\section*{Acknowledgement}
The work was supported by grants VEGA 1/0719/23, VEGA 1/0025/23, and Ministry of
Education contract No. 0466/2022.

{\setstretch{1.0}

}

\end{document}